\documentclass{article}
\usepackage{spconf,amsmath,graphicx}
\usepackage{amssymb}
\usepackage{bm}
\usepackage[subrefformat=parens]{subcaption}
\allowdisplaybreaks[4]

\newcommand{\trans}{^\mathsf{T}}
\newcommand{\hermite}{^\mathsf{H}}
\newcommand{\trace}[1]{\mathrm{tr}\left(#1\right)}
\newcommand{\const}{\mathrm{const.}}

\newcommand{\sij}{\bm{s}_{ij}}
\newcommand{\xij}{\bm{x}_{ij}}
\newcommand{\yij}{\bm{y}_{ij}}
\newcommand{\yjn}{\vec{\bm{y}}_{jn}}
\newcommand{\yjnhermite}{\yjn^{\,\mathsf{H}}}

\newcommand{\Ai}{\bm{A}_i}
\newcommand{\Wi}{\bm{W}_i}
\newcommand{\win}{\bm{w}_{in}}

\newcommand{\Rjn}{\bm{R}_{jn}}
\newcommand{\vjkn}{v_{kjn}}
\newcommand{\Ukn}{\bm{U}_{kn}}

\newcommand{\alphajn}{\alpha_{jn}}
\newcommand{\pijn}{\pi_{jn}}

\newcommand{\Qin}{\bm{Q}_{in}}
\newcommand{\gammain}{\bm{\gamma}_{in}}
\newcommand{\zetain}{\bm{\zeta}_{in}}
\newcommand{\zetahatin}{\hat{\bm{\zeta}}_{in}}
\newcommand{\etain}{\eta_{in}}
\newcommand{\etahatin}{\hat{\eta}_{in}}
\newcommand{\en}{\bm{e}_n}

\newcommand{\bin}{\bm{b}_{in}}

\newcommand{\Skn}{\bm{S}_{kn}}
\newcommand{\Tkn}{\bm{T}_{kn}}

\newcommand{\Rjntilde}{\tilde{\bm{R}}_{jn}}
\newcommand{\E}{\bm{E}}

\title{convergence-guaranteed independent positive semidefinite tensor analysis based on Student's t distribution}
\name{\shortstack{Tatsuki Kondo$^{1}$ \qquad Kanta Fukushige$^{1}$ \qquad Norihiro Takamune$^{1}$ \qquad Daichi Kitamura$^{2}$ \\ Hiroshi Saruwatari$^{1}$ \qquad Rintaro Ikeshita$^{3}$ \qquad Tomohiro Nakatani$^{3}$} \thanks{This work was partly supported by SECOM Science and Technology Foundation and JSPS KAKENHI Grant Numbers JP19H01116 and JP19K20306.}}
\address{$^{1}$ The University of Tokyo, Tokyo, Japan \\ $^{2}$ National Institute of Technology, Kagawa College, Kagawa, Japan \\ $^{3}$ NTT Communication Science Laboratories, Kyoto, Japan}

\begin{document}
\ninept

\maketitle

\begin{abstract}
In this paper, we address a blind source separation (BSS) problem and propose a new extended framework of independent positive semidefinite tensor analysis (IPSDTA).
IPSDTA is a state-of-the-art BSS method that enables us to take interfrequency correlations into account, but the generative model is limited within the multivariate Gaussian distribution and its parameter optimization algorithm does not guarantee stable convergence.
To resolve these problems, first, we propose to extend the generative model to a parametric multivariate Student's $t$ distribution that can deal with various types of signal.
Secondly, we derive a new parameter optimization algorithm that guarantees the monotonic nonincrease in the cost function, providing stable convergence.
Experimental results reveal that the cost function in the conventional IPSDTA does not display monotonically nonincreasing properties. On the other hand, the proposed method guarantees the monotonic nonincrease in the cost function and outperforms the conventional ILRMA and IPSDTA in the source-separation performance.

\end{abstract}
\section{Introduction}
\label{sec:intro}

Convolutive blind source separation (BSS)~\cite{sawada2019review} is a technique for estimating source signals from observed mixtures without any information about the mixing system, e.g., the positions of microphones and sources, or the shape of the room.
In a determined or overdetermined situation (number of microphones $\geq$ number of sources), frequency-domain independent component analysis (FDICA)~\cite{smaragdis1998blind, saruwatari2006blind}, independent vector analysis (IVA)~\cite{hiroe2006solution, kim2006independent, kim2006blind}, and independent low-rank matrix analysis (ILRMA)~\cite{kitamura2016determined, kitamura2018determined} have been proposed to solve the BSS problem. Among these methods, ILRMA provides a higher source-separation performance.
ILRMA estimates the source signals by assuming statistical independence between different sources and low-rankness in the time-frequency structure of a spectrogram represented by nonnegative matrix factorization (NMF)~\cite{lee1999learning}.

Recently, independent positive semidefinite tensor analysis (IPSDTA)~\cite{ikeshita2018independent} has been proposed. In IPSDTA, positive semidefinite tensor factorization (PSDTF)~\cite{yoshii2013infinite}, an extension of NMF, is introduced into the source model of ILRMA.
In PSDTF, we assume that the vector whose elements are the complex spectrogram of all frequency bins obeys the multivariate complex Gaussian distribution at each time frame, and that its covariance matrix is represented by a conic sum of time-invariant positive semidefinite matrices.
This modeling enables us to take interfrequency correlations into account explicitly in IPSDTA, and it is reported that IPSDTA outperforms ILRMA in the BSS task for speech.
Although the IPSDTA framework itself is a promising approach, the major drawbacks of the conventional IPSDTA are as follows: (I) The generative model is limited within the  Gaussian distribution and has less versatility.
(II) For the optimization algorithm of the conventional IPSDTA, no discussion on the convergence (the monotonic nonincrease in the cost function) has been reported.

In this paper, we provide two contributions, namely, generalization of the generative model and a new convergence-guaranteed optimization algorithm.
First, we extend the generative model of IPSDTA to a multivariate complex Student's $t$ distribution; this is hereafter referred to as $t$-{\it IPSDTA}.
Student's $t$ distribution is a parametric distribution including the Gaussian and Cauchy distributions, making $t$-IPSDTA versatile for various types of signal.
Second, we reveal that the cost function of the conventional IPSDTA does not display monotonically nonincreasing properties. To cope with this problem, we propose an optimization algorithm of $t$-IPSDTA that strictly guarantees the monotonic nonincrease in the cost function using the auxiliary function method~\cite{hunter2000quantile} and vectorwise coordinate descent (VCD)~\cite{mitsui2018vectorwise}.
Experimental results show that the proposed $t$-IPSDTA outperforms the conventional ILRMA and IPSDTA in source-separation accuracy.

\section{CONVENTIONAL METHOD}

\subsection{Formulation}

The source signal, observed signal, and separated signal in each time-frequency slot obtained via short-time Fourier transform (STFT) are denoted as
\begin{align}
\sij & =(s_{ij1}, \ldots, s_{ijn}, \ldots, s_{ijN})\trans\in\mathbb{C}^N, \\
\xij & =(x_{ij1}, \ldots, x_{ijm}, \ldots, x_{ijM})\trans\in\mathbb{C}^M, \\
\yij & =(y_{ij1}, \ldots, y_{ijn}, \ldots, y_{ijN})\trans\in\mathbb{C}^N,
\end{align}
where $\trans$ denotes the matrix transpose, and $i=1,\ldots,I,\ j=1,\ldots,J,\ n=1,\ldots,N,$ and $m=1,\ldots,M$ are indices of the frequency bins, time frames, sources, and microphones, respectively.
Assume that the mixing system is linear time-invariant and can be expressed by complex instantaneous mixing in the time-frequency domain. Under this condition, the observed signal can be represented as $\xij=\Ai\sij$, where $\Ai\in\mathbb{C}^{M\times N}$ is a time-invariant mixing matrix for each frequency bin.
If $N=M$ and $\Ai$ is invertible, a demixing matrix $\Wi=(\bm{w}_{i1},\ldots,\win,\ldots,\bm{w}_{iN})\hermite=\Ai^{-1}$ exists, and the separated signal can be estimated as
\begin{equation}
\yij=\Wi\xij,
\end{equation}
where $\hermite$ denotes the Hermitian transpose.

\vspace{-6pt}
\subsection{Conventional IPSDTA}
\vspace{-4pt}

\begin{figure}[tb]
\begin{center}
\includegraphics[width=\columnwidth]{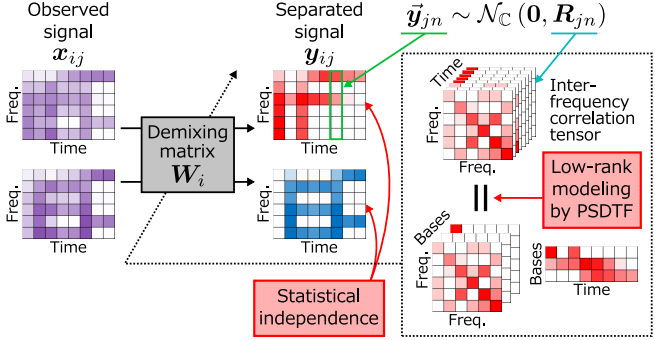}
\vspace{-25pt}     
\end{center}
\caption{Overview of IPSDTA.}
\label{fig:IPSDTA}
\vspace{-10pt}   
\end{figure}

In the conventional IPSDTA~\cite{ikeshita2018independent}, the vector whose elements are the complex spectrogram of all frequency bins is assumed to follow the multivariate complex Gaussian distribution (see Fig.\,\ref{fig:IPSDTA}).
The task of the conventional IPSDTA is to find the parameters to minimize the cost function, the negative log-likelihood of the observed signal, under the assumption of independence between sources.
The algorithm alternately updates the demixing matrix $\Wi$ and the source model.
The demixing matrix is updated by an {\it interfrequency-correlation-aware} algorithm extended from the iterative projection (IP); IP itself is a fast and convergence-guaranteed algorithm used in FDICA~\cite{ono2010auxiliary}, IVA~\cite{ono2011stable}, and ILRMA~\cite{kitamura2016determined} that does not take interfrequency correlations into account.
The source model is updated by the expectation-maximization algorithm as in PSDTF~\cite{liutkus2017diagonal}.

In the algorithm that updates the demixing matrix, it is difficult to solve an equation to find the stationary point of the cost function in a closed-form manner. Instead, a fixed-point iteration is introduced to solve the equation.
Since the fixed-point iteration does not always guarantee the convergence in general, the conventional IPSDTA suffers from the lack of stability in parameter optimization and source separation;
this will be experimentally shown in Sec.\,\ref{sec:result_convergence}.

\vspace{-3pt}
\section{proposed probabilistic model}
\label{sec:proposed}
\vspace{-4pt}

In this section, the generative model of the conventional IPSDTA is extended to the multivariate complex Student's $t$ distribution.
By assuming the independence of the separated signal, we can model the probability distribution of the observed signal as
\begin{equation}
p\left(\left\{\xij\right\}_{i,j}\right)=\prod_{n}p\left(\left\{\yjn\right\}_{j}\right)\cdot\prod_{i,j}\left|\det\Wi\right|^2,
\end{equation}
where $\{z_l\}_l$ denotes a set whose elements are $z_l$ for all $l$.
In the proposed $t$-IPSDTA, we assume the following two conditions for the generative model of $\yjn$ $=(y_{1jn}, \ldots, y_{ijn}, \ldots, y_{Ijn})\trans\in\mathbb{C}^I$.
\begin{itemize}
\item[(i)]
$\yjn$ obeys the multivariate complex Student's $t$ distribution $\mathcal{T}_{\nu}\left(\bm{0},\Rjn\right)$ for each $j$ and $n$ independently. The $d$-dimensional complex Student's $t$ distribution $\mathcal{T}_{\nu}\left(\bm{\mu},\bm{\Sigma}\right)$ is defined by the following probability density function:
\begin{align}
p(\bm{z};\bm{\mu},\bm{\Sigma}) & =\frac{2^d}{(\nu\pi)^d}\frac{\Gamma(\frac{\nu+2d}{2})}{\Gamma(\frac{\nu}{2})}\frac{1}{\det\bm{\Sigma}} \nonumber \\
& \quad\times\!\left(1\!+\!\frac{2}{\nu}(\bm{z}\!-\!\bm{\mu})\hermite\bm{\Sigma}^{-1}(\bm{z}\!-\!\bm{\mu})\right)^{-\frac{\nu+2d}{2}},
\end{align}
where $\nu$ is the degree-of-freedom parameter of the Student's $t$ distribution. This distribution corresponds to the multivariate complex Cauchy distribution when $\nu=1$ and to the multivariate complex Gaussian distribution when $\nu\rightarrow\infty$. $\bm{\mu}\in\mathbb{C}^d$ and $\bm{\Sigma}\in\mathbb{C}^{d\times d}$ are the parameters corresponding to an average vector and a covariance matrix in the multivariate complex Gaussian distribution, respectively.
\item[(ii)]
For the $n$th source, Hermitian positive semidefinite matrices $\{\Rjn\}_j$ are modeled by PSDTF~\cite{yoshii2013infinite} as follows:
\begin{equation}
\label{eq:PSDTF}
\Rjn=\sum_{k=1}^{K_n}\vjkn\Ukn,
\end{equation}
where $K_n$ is the number of bases in PSDTF, $\vjkn\geq0$ is a time-variant activation, and $\Ukn\in\mathbb{C}^{I\times I}$ is a time-invariant Hermitian positive semidefinite matrix.
\end{itemize}
By assuming conditions (i) and (ii), we can obtain the negative log-likelihood as
\begin{align}
\mathcal{L} & = \sum_{j,n}\left(\log\det\Rjn+\frac{\nu+2I}{2}\log\left(1+\frac{2}{\nu}\yjnhermite\Rjn^{-1}\yjn\right)\right) \nonumber \\
\label{eq:L}
& \quad-J\sum_i\log|\det\Wi|^2+\const,
\end{align}
where the $\const$ term does not depend on the parameters.
The aim of the $t$-IPSDTA algorithm is to minimize the cost function $\mathcal{L}$ with respect to $\Wi, \vjkn$, and $\Ukn$.
When $\nu\rightarrow\infty$, the generative model of $t$-IPSDTA coincides with that of the conventional IPSDTA, and by changing $\nu$, we can deal with various types of signal.

\vspace{-1pt}
\section{proposed optimization algorithm}
\vspace{-1pt}

\subsection{Algorithm overview}
\label{sec:algorithm_overview}
\vspace{-1pt}

In this section, a new algorithm for minimizing the cost function $\mathcal{L}$ is presented.
It is based on the auxiliary function method~\cite{hunter2000quantile} and a different type of coordinate descent from IP, and consequently guarantees the monotonic nonincrease in the cost function, whereas the conventional IPSDTA does not guarantee these properties.
The second and third terms in (\ref{eq:L}) are related to the demixing matrix $\Wi$ (or $\win$) because $\yjn$ includes $y_{ijn}=\win\hermite\xij$, and the first and second terms are related to the source model $\vjkn$ and $\Ukn$. The algorithm alternately updates the demixing matrix $\Wi$ and the source model $\vjkn$ and $\Ukn$; each update rule is derived in the following subsections.

\vspace{-2pt}
\subsection{Update of demixing matrix $\Wi$}
\vspace{-1pt}

When we describe only the terms related to the demixing matrix, the cost function $\mathcal{L}$ is described as
\begin{align}
\mathcal{L} & = \sum_{j,n}\frac{\nu+2I}{2}\log\left(1+\frac{2}{\nu}\yjnhermite\Rjn^{-1}\yjn\right) \nonumber \\
\label{eq:L12}
& \quad -J\sum_i\log|\det\Wi|^2+\const
\end{align}
In general, since $\log z$ is a concave function on $z>0$, the following inequality holds:
\begin{equation}
\label{eq:logx}
\log z\leq\frac{1}{c}\left(z-c\right)+\log c,\quad c>0.
\end{equation}
The equality of (\ref{eq:logx}) holds if and only if $c=z$.
By applying (\ref{eq:logx}) to (\ref{eq:L12}), we can design the auxiliary function $\mathcal{L}^+$ as
\begin{align}
\mathcal{L} & \leq \sum_{j,n}\frac{\nu+2I}{2}\left(\frac{1}{\alphajn}\left(1+\frac{2}{\nu}\yjnhermite\Rjn^{-1}\yjn-\alphajn\right)+\log\alphajn\right) \nonumber \\
& \quad-J\sum_i\log|\det\Wi|^2+\const \nonumber \\
& =\sum_{j,n}\pijn\yjnhermite\Rjn^{-1}\yjn-J\sum_i\log|\det\Wi|^2+\const \nonumber \\
\label{eq:L+}
& \equiv\mathcal{L}^+,
\end{align}
where $\alphajn$ is an auxiliary variable and $\pijn=(\nu+2I)/(\nu\alphajn)$.
The equality of (\ref{eq:L+}) holds if and only if
\begin{equation}
\alphajn=1+\frac{2}{\nu}\yjnhermite\Rjn^{-1}\yjn,
\end{equation}
i.e.,
\begin{equation}
\label{eq:pi}
\pijn=\frac{\nu+2I}{\nu+2\yjnhermite\Rjn^{-1}\yjn}.
\end{equation}
When we describe only the terms related to the demixing matrix of the $i$th frequency bin, $\Wi$, and regard $\bm{W}_{i'} (i'\neq i)$ as a constant, the auxiliary function $\mathcal{L}^+$ is expanded as
\begin{align}
\frac{1}{J}\mathcal{L}^+ & = \sum_n\left(\win\hermite\Qin\win+\win\hermite\gammain+\gammain\hermite\win\right)\nonumber \\
\label{eq:L/J}
& \quad-\log|\det\Wi|^2+\const,
\end{align}
where
\begin{align}
\label{eq:Q}
\Qin & =\frac{1}{J}\sum_j\left[\left(\pijn^{-1}\Rjn\right)^{-1}\right]_{ii}\xij\xij\hermite, \\
\label{eq:gamma}
\gammain & =\sum_{i'\neq i}\left(\frac{1}{J}\sum_j\left[\left(\pijn^{-1}\Rjn\right)^{-1}\right]_{i'i}\xij\bm{x}_{i'j}\hermite\bm{w}_{i'n}\right).
\end{align}
Here, $[(\pijn^{-1}\Rjn)^{-1}]_{i'i}$ denotes the $(i',i)$th element of the matrix $(\pijn^{-1}\Rjn)^{-1}$.
Equation\,(\ref{eq:L/J}) is the sum of the log-determinant of $\Wi$, the quadratic form of $\win$, and {\it linear terms of} $\win$.
This type of problem cannot be solved by IP because of the existence of the linear terms.
VCD, which we previously proposed~\cite{mitsui2018vectorwise}, is an optimization algorithm that can be applied to a cost function of this form.
In (\ref{eq:L/J}), $|\det\Wi|^2$ is rewritten as $|\win\hermite\bin|^2$, where $\bin$ is the $n$th column of the cofactor matrix of $\Wi$.
Since $\bin$ is independent of $\win$ owing to the definition of the cofactor matrix~\cite{strang1993introduction}, the partial derivative of $\mathcal{L}^+/J$ with respect to $\win^*$ is obtained as
\begin{align}
\frac{\partial}{\partial\win^*}\left(\frac{1}{J}\mathcal{L}^+\right) = \Qin\win+\gammain-\frac{\bin}{\win\hermite\bin},
\end{align}
where $^*$ denotes complex conjugate.
By solving the equation $\partial(\mathcal{L}^+/J)/\partial\win^*=0$, we describe the update rules of $\win$ based on VCD as follows:
\begin{align}
\label{eq:zeta}
\zetain & \leftarrow(\Wi\Qin)^{-1}\en, \\
\zetahatin & \leftarrow\Qin^{-1}\gammain, \\
\etain & \leftarrow\zetain\hermite\Qin\zetain, \\
\etahatin & \leftarrow\zetain\hermite\Qin\zetahatin, \\
\label{eq:w}
\win & \leftarrow
\begin{cases}
\frac{\zetain}{\sqrt{\etain}}-\zetahatin & (\etahatin=0) \\
\frac{\etahatin}{2\etain}\left(1-\sqrt{1+\frac{4\etain}{|\etahatin|^2}}\right)\zetain-\zetahatin & (\mathrm{otherwise}),
\end{cases}
\end{align}
where it has been proved that these update rules can minimize $\mathcal{L}^+$ with respect to $\win$~\cite{mitsui2018vectorwise}. Therefore, the demixing matrix $\Wi$ can be estimated by iteratively updating the parameters via (\ref{eq:pi}), (\ref{eq:Q}), (\ref{eq:gamma}), and (\ref{eq:zeta})--(\ref{eq:w}).

\subsection{Update of source model $\vjkn$ and $\Ukn$}

When we describe only the terms related to the source model, the cost function $\mathcal{L}$ is described as
\begin{align}
\mathcal{L} & =\sum_{j,n}\left(\log\det\Rjn+\frac{\nu+2I}{2}\log\left(1+\frac{2}{\nu}\yjnhermite\Rjn^{-1}\yjn\right)\right) \nonumber \\
\label{eq:cost_tIPSDTA}
& \quad+\const
\end{align}
This is equivalent to the model of $t$-PSDTF~\cite{yoshii2016student} with the observation $\yjn$. By applying the $t$-PSDTF algorithm to (\ref{eq:cost_tIPSDTA}), we derive the update rules as follows:
\begin{align}
\pijn & =\frac{\nu+2I}{\nu+2\yjnhermite\Rjn^{-1}\yjn}, \\
\label{eq:Skn}
\Skn & =\sum_j\vjkn\Rjn^{-1}\left(\pijn\yjn\yjnhermite\right)\Rjn^{-1}, \\
\Tkn & =\sum_j\vjkn\Rjn^{-1}, \\
\label{eq:vjkn}
\vjkn & \leftarrow\vjkn\sqrt{\frac{\trace{\pijn\yjn\yjnhermite\Rjn^{-1}\Ukn\Rjn^{-1}}}{\trace{\Rjn^{-1}\Ukn}}}, \\
\Ukn & \leftarrow\Ukn\Skn^{\frac{1}{2}}\left(\Skn^{\frac{1}{2}}\Ukn\Tkn\Ukn\Skn^{\frac{1}{2}}\right)^{-\frac{1}{2}}\Skn^{\frac{1}{2}}\Ukn.
\end{align}
To avoid the ambiguity of the scales of $\vjkn$ and $\Ukn$, we adjust the scales at each iteration so that $\trace{\Ukn}=1$.

\subsection{Interpretation on update rule of source model}
\label{sec:interpretation}

When $\nu\rightarrow\infty$ (multivariate complex Gaussian distribution), $\pijn\yjn\yjnhermite$ becomes $\yjn\yjnhermite$. Hence, $\pijn\yjn\yjnhermite$ in (\ref{eq:Skn}) and (\ref{eq:vjkn}) can be interpreted as a virtual instantaneous covariance matrix of the separated signal.
Then, the following equation holds:
\begin{align}
& \pijn\yjn\yjnhermite=\yjn\left(\lambda+(1-\lambda)\yjnhermite\Rjntilde^{-1}\yjn\right)^{-1}\yjnhermite,
\end{align}
where $\lambda=\nu/(\nu+2I), \Rjntilde=I\Rjn$. By applying the matrix inversion lemma to $(\lambda+(1-\lambda)\yjnhermite\Rjntilde^{-1}\yjn)^{-1}$, we have
\begin{align}
\label{eq:tIPSDTA_harmonic_mean}
& \pijn\yjn\yjnhermite \nonumber \\
&\,= \Rjntilde\left(\lambda\Rjntilde+(1-\lambda)\yjn\yjnhermite\right)^{-1}\yjn\yjnhermite
\nonumber \\
&\,= \lim_{\varepsilon\rightarrow0}\Rjntilde\left(\lambda\Rjntilde+(1-\lambda)\left(\yjn\yjnhermite+\varepsilon\E\right)\right)^{-1}\left(\yjn\yjnhermite+\varepsilon\E\right) \nonumber \\
&\,= \lim_{\varepsilon\rightarrow0}\left(\lambda\left(\yjn\yjnhermite+\varepsilon\E\right)^{-1}+(1-\lambda)\Rjntilde^{-1}\right)^{-1},
\end{align}
\vspace{-24pt}

\noindent
where $\E$ is the identity matrix.
When $\varepsilon\rightarrow0$, $\yjn\yjnhermite+\varepsilon\E$ becomes $\yjn\yjnhermite$.
Thus, the virtual instantaneous covariance matrix $\pijn\yjn\yjnhermite$ can be interpreted as the harmonic mean of the real instantaneous covariance matrix $\yjn\yjnhermite$ and the source model obtained in the previous iteration, $\Rjntilde=I\Rjn$, with a ratio of $\nu:2I$.
As $\nu$ becomes smaller, $\vjkn$ and $\Ukn$ are updated taking $\Rjntilde$ into account more strictly.
This implies that we can avoid the overfitting for the temporarily separated $\yjn$ and maintain the low-rankness of the source model.

Note that, regarding the limited case for scalar variables in Student's $t$-distribution-based NMF~\cite{yoshii2016student}, such harmonic mean properties have been indicated.
On the other hand, our derived {\it matrix harmonic mean} formulation is the world's first interpretation and mathematical generalization for the multivariate case to the best of our knowledge.

\vspace{-4pt}
\section{EXPERIMENT}
\vspace{-1pt}

\subsection{Experimental conditions}

We conducted a two-source separation experiment using the SiSEC2008 dataset~\cite{vincent20092008} (No.\,1 and No.\,3 in dev1\_male4, No.\,2 and No.\,4 in dev1\_male4, No.\,1 and No.\,3 in dev1\_female4, or No.\,2 and No.\,4 in dev1\_female4). 
We compared three methods, namely ILRMA, the conventional IPSDTA, and the proposed $t$-IPSDTA.
In the proposed $t$-IPSDTA, VCD was performed 10 times at each PSDTF update.
The initial values of $\Wi$, $\vjkn$, and $\Ukn$ were set to the identity matrix, a random number with a uniform distribution over $(0, 1)$, and a matrix that includes random numbers with a uniform distribution over $(0, 1)$ in the diagonal entries, respectively.
Similarly to \cite{ikeshita2018independent}, we divided the set of frequency bins, $\{1,\ldots,2049\}$, into $E_1=\{1,2\},\ldots,E_{1023}=\{2045,2046\}$, and $E_{1024}=\{2047,2048,2049\}$, and imposed the block decomposition on $\Ukn$ via $\{E_l\}_l$.
The sampling frequency was 16\,kHz and STFT was carried out using a 256-ms-long Hamming window with a 128\,ms shift.
The total number of iterations was 100.
The interelement spacing $\delta$ was set to 5\,cm or 1\,m, and the reverberation time (RT) was set to 130\,ms or 250\,ms.
The evaluation score was the source-to-distortion ratio (SDR) improvement~\cite{vincent2006performance}.

\subsection{Results for convergence behavior}
\label{sec:result_convergence}

\begin{figure}[tb]
\begin{center}
\includegraphics[width=\columnwidth]{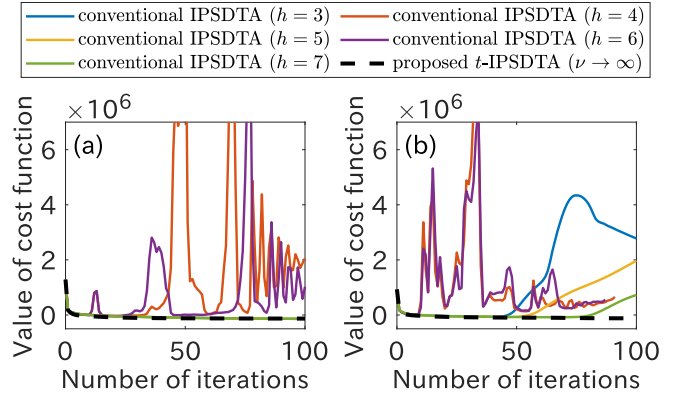}
\vspace{-25pt}     
\end{center}
\caption{Behavior of cost function: (a) male pair, RT=130\,ms, $\delta$=5\,cm, and (b) female pair, RT=250\,ms, $\delta$=5\,cm. $h$ is number of repeated steps in fixed-point iteration in conventional IPSDTA~\cite{ikeshita2018independent}.}
\label{fig:cost}
\end{figure}

Fig.\,\ref{fig:cost} shows the values of cost functions for the conventional IPSDTA and the proposed $t$-IPSDTA with $\nu\rightarrow\infty$, where their generative models are the same.
In this figure, we omit the term $\const$ in (\ref{eq:L}).
The number of bases was set to two. 

For the conventional IPSDTA, the values of cost functions increase in the middle of the 100 iterations, especially at $h=4,6$, showing no guarantee of monotonically nonincreasing properties.
On the other hand, for the proposed $t$-IPSDTA, the value of the cost function monotonically decreases, which is consistent with the properties described in Sec.\,\ref{sec:algorithm_overview}.
Furthermore, the convergence speed of the proposed algorithm is the same as that of the conventional IPSDTA.
From these results, the advantage of the proposed $t$-IPSDTA is revealed in terms of the convergence behavior.

\subsection{Results for separation performance}
\label{sec:result_performance}

\begin{figure}[tb]
\begin{center}
\includegraphics[width=\columnwidth]{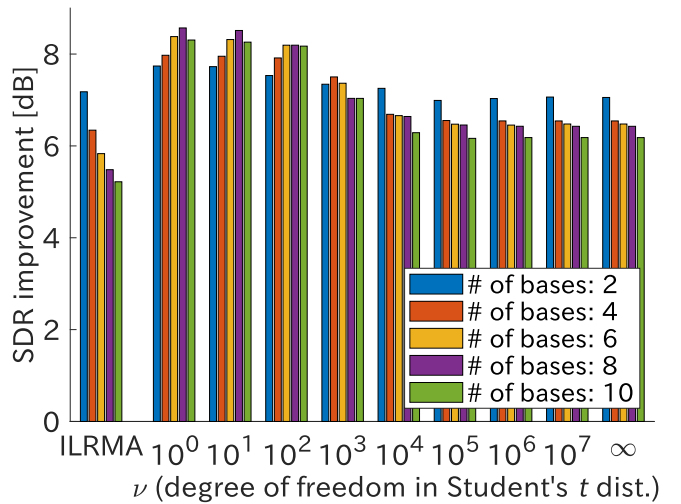}
\vspace{-25pt}     
\end{center}
\caption{SDR improvements for each method, where $\nu\rightarrow\infty$ corresponds to multivariate complex Gaussian distribution used in \cite{ikeshita2018independent}.}
\label{fig:SDR}
\vspace{-5pt}   
\end{figure}

Fig.\,\ref{fig:SDR} shows the average SDR improvements of ILRMA and the proposed $t$-IPSDTA over the combination of speakers, the RTs, $\delta$, and 10-trial initial values of $\vjkn$ and $\Ukn$.
The number of bases, $K_n$, is changed from two to ten.
ILRMA provides the peak SDR improvement when $K_n=2$, which is the same tendency as shown in \cite{kitamura2016determined}.
The proposed $t$-IPSDTA shows the best SDR improvement when $K_n=8$ and $\nu=10^0$, which outperforms ILRMA.
As the degree-of-freedom parameter $\nu$ becomes smaller, the proposed $t$-IPSDTA shows higher separation performance.
This indicates the effectiveness of introducing the multivariate complex Student's $t$ distribution in the IPSDTA framework, i.e., 
model versatility described in Sec.\,\ref{sec:proposed} and overfitting avoidance described in Sec.\,\ref{sec:interpretation}.

\section{CONCLUSION}

In this paper, we proposed to extend the generative model of IPSDTA to the multivariate complex Student's $t$ distribution.
In addition, we derived a new parameter optimization algorithm that guarantees the monotonic nonincrease in the cost function, which the conventional IPSDTA does not guarantee in theory.
Experimental results revealed that the values of cost functions in the conventional IPSDTA do not display monotonically nonincreasing properties.
On the other hand, the proposed $t$-IPSDTA guaranteed the monotonic nonincrease in the cost function and outperformed the conventional ILRMA and IPSDTA in the SDR improvement.

%

\vfill
\pagebreak

\bibliographystyle{IEEEbib}
\bibliography{strings,refs}

\end{document}